\documentclass[prb,aps,twocolumn,showpacs,10pt]{revtex4-1}
\usepackage{graphicx}  % needed for figures
\usepackage{dcolumn}   % needed for some tables
\usepackage{bm}        % for math
\usepackage{amssymb}   % for math
\usepackage{amsmath}
\usepackage{graphicx}
\usepackage{slashed}

\newcommand{\be}{\begin{equation}}
\newcommand{\ee}{\end{equation}}
\newcommand{\bea}{\begin{eqnarray}}
\newcommand{\eea}{\end{eqnarray}}
\newcommand{\ot}{\otimes}

\newcommand{\p}{\partial}
\newcommand{\lam}{\lambda}
\newcommand{\al}{\alpha}
\newcommand{\ep}{\epsilon}
\newcommand{\s}{\sigma}

\newcommand{\la}{\langle}
\newcommand{\ra}{\rangle}
\newcommand{\lb}{\left[}
\newcommand{\rb}{\right]}
\newcommand{\lp}{\left(}
\newcommand{\rp}{\right)}

\renewcommand{\dag}{\dagger}
\renewcommand{\exp}{{\rm exp}}
\renewcommand{\cos}{{\rm \, cos\,}}

\renewcommand{\vec}[1]{{\bf #1}}

\def\nn{\nonumber\\}

\begin{document}

\title{Fractionalized gapless quantum vortex liquids}
\author{Chong Wang}
\author{T. Senthil}
\affiliation{Department of Physics, Massachusetts Institute of Technology, Cambridge, MA 02139, U. S. A.}
\date{\today}

\begin{abstract}
The standard theoretical approach to gapless spin liquid phases of two-dimensional frustrated quantum antiferromagnets invokes the concept of fermionic slave particles into which the spin fractionalizes. As an alternate we explore new kinds of gapless spin liquid phases in frustrated quantum magnets with $XY$ anisotropy where the vortex of the spin fractionalizes into gapless itinerant fermions. The resulting gapless fractionalized vortex liquid phases are studied within a slave particle framework that is dual to the usual one. We demonstrate the stability of some such phases and describe their properties.   We give an explicit construction  in an $XY$-spin-1 system on triangular lattice, and interpret it as
a critical phase in the vicinity of spin-nematic states.

\end{abstract}

\maketitle

\section{Introduction}

Quantum spin liquids are exotic phases of matter beyond 
Landau's paradigm of symmetry-breaking\cite{wen}.  In contrast to other familiar ground states of quantum magnets (such as antiferromagnets or ferromagnets) the quantum spin liquid ground state has a non-local entanglement between its local degrees of freedom.  Similar `long range entanglement' also appears in the ground state of some other states of matter, for instance in the fractional quantum Hall states, and in Fermi/non-Fermi liquid metals. Since the original conception of the possibility of the quantum spin liquid, there has been tremendous progress in describing them theoretically.  Many different kinds of quantum spin liquids are known to be theoretically possible. In the last decade a number of experimental candidates have also appeared.  Interestingly all the existing experimental candidates seem to have gapless excitations which are not related to Goldstone modes of any broken symmetry. The theory of such gapless quantum spin liquids is however much less developed than the theory of gapped 
quantum spin liquid states. 

The currently known experimental candidate spin liquid materials may be conveniently grouped into two broad categories. The first - dubbed ``weak Mott insulators" - are close to the Mott transition and have significant virtual charge fluctuations.  Both the layered organics
$\kappa-(ET)_2Cu_2(CN)_3$ and $EtMe_3 Sb[Pd(dmit)_2]_2$, and the three dimensional hyperkagome iridate $Na_4Ir_3O_8$ are all Mott insulating at ambient pressure but can be driven\cite{kanoda1,kato,takagi} through the Mott transition with application of moderate pressure.     Indeed quantum spin liquid behavior may well be a common fate of {\em weak} Mott insulators.
The second category - dubbed ``strong Mott insulators" - have large charge gaps that are well separated from their exchange scales.   These two classes of spin liquids likely require different theoretical approaches.

In weak Mott insulators gapless spin excitations are perhaps expected. At short length/time scales such insulators look roughly the same as a metal.  As confirmed by various theoretical calculations\cite{sfs1,sfs2,matthewnumerics}, it is then reasonable that at longer length scales even though the charge localizes the spin continues to be carried by itinerant neutral fermions (the spinons). 
Remarkably gapless excitations are found even in candidate spin liquids which are strong Mott insulators. Striking examples are the Kagome systems\cite{yslee,hiroi} $ZnCu_3(OH)_6Cl_2$ 
(Herbertsmithite) and $Cu_3V_2O_7(OH)_2.2H_2O$ (Volborthite).   Similarly the recently reported spin-1 spin liquid\cite{balicas} $Ba_3NiSb_2O_9$ is also a strong Mott insulator.

Recent progress in Density Matrix Renormalization Group calculations of the isotropic spin-$1/2$ Kagome magnet\cite{dmrg} reveal a large spin gap ($0.14 J)$ which is not seen in the experiments on Herbertsmithite\cite{han}. The real model for this material is more complicated and must include Dzyaloshinski-Moriya as well as other anisotropies. Further there are significant impurity effects attributed to excess Cu spins sitting in between the Kagome planes.  Other complications may exist in other materials. Nevertheless the surprisingly common occurance of gapless spin liquids in strongly Mott insulating materials leads to some fundamental questions in the theory of spin liquids.
 
In what theoretical framework should we discuss these gapless spin liquids? 
Currently one framework that is known is to start with a slave particle description of the
physical spin operator in terms of fermionic neutral spin-$1/2$
spinons. The resulting spinon Hamiltonian is then first treated at
a mean field level. At this level of treatment the spinon spectrum
may well be gapless (with Fermi points or even a Fermi surface).
Going beyond mean field requires including fluctuations. The
resulting theory typically includes a fluctuating gauge field.
Thus in this approach gapless spin liquids are described by an
effective theory that involves gapless fermionic spinons coupled
to a fluctuating gauge field. If this theory is stable  then this
is a legitimate description of a possible gapless spin liquid
phase. 

The slave particle approach described above is deservedly popular and it certainly enables description of a class of quantum spin liquids.  
However while this seems natural for weak Mott insulators (as is confirmed by many existing calculations) it is hardly obvious that this is the way forward in dealing with gapless spin liquids in, say, the Kagome magnets, or in the spin-$1$ magnet. As currently no other methods are known, fermionic spinon based approaches are the ``knee-jerk" reaction of theorists to the announcement of any experimental candidate gapless spin liquid. 
A big open question in the field is whether there are other approaches that enables access to a different class of gapless spin liquids.  More specifically do gapless spin liquids exist that are beyond the existing fermionic spinon (+ gauge field) paradigm? If so what is their phenomenology? Could they be more natural candidates for some of the 
spin liquids that are reported?  

In this paper we introduce a new theoretical route to gapless quantum spin liquids in spin systems with XY symmetry  which appears to be distinct from the conventional fermionic spinon route.  We utilize a dual description of such a spin system in terms of vortices in the XY spin.  We show that quantum vortex liquid phases exist where there are gapless fermionic fields that carry the vorticity. We access these gapless vortex liquid phases through a `dual' parton approach where we fractionalize the fundamental vortex field into fermionic half-vortices. These may then form a gapless state. We describe an example of this construction for spin-1 quantum XY models on a honeycomb lattice. The dual parton approach is complementary to the standard one which fractionalizes the physical spin itself. Indeed it is likely that the phases we access have no simple and intuitive description within the standard approach. In Sec. \ref{parton} we give a construction of the phases in the usual parton language, but the 
construction is quite complicated and requires auxiliary degrees of freedom on the lattice. This makes it much more natural to think in terms of fractionalizing vortices rather than spins.

An interesting earlier attempt with a motivation similar to ours was made in Ref. \onlinecite{avl1,avl2,avl3}, where the vortices were ``fermionized'' through a Chern-Simons flux-attachment. The fermions can then be put into a gapless band, and the resulting state becomes gapless and $U(1)$-symmetric. 
However,  this construction has problems with implementing time-reversal symmetry.   In the simplest context in which such a fermionized vortex duality was attempted, it was shown in Ref. \onlinecite{topo} that such an approach would require an extra topological term into the original (un-dualized) description. Moreover, it was found recently\cite{avts12,hmodl} that such states realize time-reversal symmetry anomalously, and could 
appear only on the surface of certain bosonic topological insulators. Therefore such a state realized in two dimensions will break time-reversal symmetry (an example was discussed in Ref.~\onlinecite{2dvrtxmtl}), and hence is not suitable to describe symmetric quantum spin liquids. 

Closer to our approach is  Ref. \onlinecite{hermele} which also employed a dual fermionic parton decomposition of the fundamental vortex field. The goal however was different from ours and that work did not attempt to find stable quantum spin liquid phases through the dual parton approach.

\section{Duality, vortices, and fractionalization}
\label{duality}

A spin system with $XY$ symmetry can be fruitfully viewed as a system of interacting bosons (with $S^z$ playing the role of boson number and $S^+$ the role of the boson creation operator $b^\dagger$). For a bosonic system with global~$U(1)$ symmetry, it is known that one can make a 
duality mapping and describe the system in terms of vortices\cite{dual}. Specifically, one can write the conserved $U(1)$ current as the flux of a 
non-compact $U(1)$ gauge field
\begin{equation}
 j^{\mu}=\frac{\epsilon^{\mu\nu\lambda}}{2\pi}\partial_{\nu}a_{\lambda},
\end{equation}
The gauge field $a_{\mu}$ couples to a formally bosonic field $\Phi$ that corresponds to vortices in the order parameter of the global $U(1)$ symmetry. 
If the vortices are gapped, we get a superfluid/ordered magnet with the global $U(1)$ symmetry broken, in which the gapless photons of the $a_{\mu}$ gauge field 
corresponds to the Goldstone mode. But if the vortices are condensed instead, the whole system will be gapped due to the Higgs mechanism and we get a trivial Mott insulator/paramagnet. One can then ask the following question: 
is it possible for the vortices to be in a stable gapless phase, so that the whole system is gapless while the global $U(1)$ symmetry is still preserved?

 The route we will take is to fractionalize the vortex into two fermions, schemetically we have
\begin{equation}
\label{vfrac}
 \Phi\sim\psi_1\psi_2,
\end{equation}
where $\Phi$ represents the vortex field rather than the physical spin as in usual parton construction, and $\psi_{1,2}$ are fermions representing ``fractionalized'' vortices. Such a ``dual'' parton construction can easily be made time-reversal invariant. 

As in the usual parton construction the dual parton representation introduces an $SU(2)$ gauge redundancy. In this paper we will restrict ourselves to states where this $SU(2)$ gauge structure is broken down to $Z_2$.  This will already be enough to produce a number of interesting states of the spin/boson system. 

Before describing the gapless states we are interested in let us briefly describe some conventional states that will help build intuition about these fractionalized vortices. Consider the simplest such fractionalized vortex state, in which the fermionic fractional vortices $\psi$ is gapped, and couple to $a_\mu$ with gauge charge $1/2$. Then we may integrate them out to get a Maxwell action for the $a_\mu$. The gauge field fluctuations are thus gapless. 

Physically this is a superfluid phase of the original bosons. However the presence of the gapped fractional vortex means that it is a {\em paired} superfluid where boson pairs $b^2$ are condensed without condensation of individual bosons $b$. (In spin language this is a `spin nematic' phase).  The excitation spectrum of such a paired superfluid is well known. There is the usual gapless superfluid sound mode which in the dual description is identified with the propagating photon. The single boson survives as  a gapped `Bogoliubov' quasiparticle, and may be described as an Ising spin $s$. In addition there is a half-vortex excitation where the phase of $b^2$ winds by $2\pi$. The Ising spin $s$ in turn acquires a phase $\pi$ upon encircling this vortex. Thus the Ising spin and  the half-vortex are mutual semions.   If we assign bose statistics to the half-vortex, its bound state with the Ising spin $s$ yields an excitation that is a fermion and also carries half-vorticity. Clearly we identify this with the 
$\psi$ particles in the dual parton description.  

Since we have assumed a state that has broken the dual $SU(2)$ gauge structure to $Z_2$, the $\psi$ carry a $Z_2$ gauge charge (in addition to the $U(1)$ gauge charge representing their vorticity). Correspondingly there is a $Z_2$ gauge vortex (the vision) which clearly must be identified with the $s$ particle, {\em i.e} the unpaired boson in the paired superfluid.

The original physical boson is the composite of a vison $s$ and 
a $2\pi$-flux of the U(1) gauge field. Condensing the original boson means condensing the vison $s$, which confines the half-vortices, in agreement 
with the usual description.

%The above interpretation can be made more precise by writing the original boson using a slave Ising-spin:
%\be
%b=sb',
%\ee
%where $s=\pm 1$ and both $s$ and $b'$ couple to a $\mathbb{Z}_2$ gauge field, hence forming mutual semions with the $\mathbb{Z}_2$ flux $v$. We assume that $v$ is gapped so that the theory is deconfined. One can then perform the usual duality transform on $b'$, and get a boson (vortex) coupled with a non-compact U(1) gauge field $a_{\mu}$. But notice here that the gapped $\mathbb{Z}_2$ flux $v$ and $b'$ form a pair of mutual semions, so the $b'$ boson sees $v$ as a "half vortex", hence the vison should couple to the U(1) gauge field in the dual picture minimally with half charge. Similarly the bound states of $v$ and $s$ (which are fermions denoted as $\psi\sim vs$) also should couple with the gauge field with half charges. If everything is gapped except the gauge field $a_{\mu}$, then clearly $\la b'\ra \neq 0$ and hence $b\sim s$. If the Ising spin $s$ is disordered, then we have $\la b^2\ra \sim \la s^2\ra \neq 0$ but $\la b\ra \sim\la s\ra =0$, which gives a nematic state. If $s$ is ordered instead, we 
%have $\la b\ra \neq 0$ which is a simple ordered state.

One can also consider a different phase in which the $\psi$ fermions are paired $\la \psi\psi\ra \neq0$. In such a phase $a_{\mu}$ is gapped, and we get a  fractionalized liquid 
with $Z_2$ topological order.  The pair condensation quantizes the magnetic flux of $a_\mu$ in units of $2\pi$, which corresponds to an excitation $b_v$ with physical charge $1$ and boson statistics.  This $b_v$ is however not to be identified with the physical boson $b$. Indeed the unpaired $\psi$ fermion survives as a Bogoliubov quasiparticle which is a mutual semion with the $b_v$.  This is in contrast with the physical boson $b$ which is local with respect to all excitations. The state obtained this way has the topological order of a $Z_2$ quantum spin liquid but with symmetry realized in an unfractionalized manner.

The most interesting situation - which we explore in this paper -  is when we put the $\psi$ fermions into a gapless band structure, such as a massless Dirac band.  The gapless fermions will then couple to the 
gauge field $a_{\mu}$ strongly, and form a gapless state which is not ordered.  This is a gapless quantum spin liquid state which is potentially not accessible within the standard fermionic spinon-gauge field paradigm.

\section{Construction with frustrated quantum $XY$ model}
We now illustrate the construction of an example of such a gapless fractionalized quantum vortex liquid. 
Consider a quantum $XY$ antiferromagnet on a two-dimensional triangular lattice. The Hamiltonian can be written as a rotor model ($b\sim e^{i\phi}$) in a 
background static gauge field $\mathcal{A}^0$:
\be
\label{rotor}
H=-J\sum_{\la ij\ra }\cos\lp\phi_i-\phi_j+\mathcal{A}^0_{ij} \rp +U\sum_i n_i^2+...
\ee
where $\mathcal{A}^0$ gives a $\pi$ flux on each triangular plaquette (corresponding to antiferromagnetic exchange).  We can think of the $\pi$ flux as requiring that there be an average vortex filling of $1/2$ per site on the dual honeycomb lattice. Going then to the vortex picture, we get a theory 
of hard-core bosons (the vortices) at half-filling on the honeycomb lattice, coupled with a non-compact U(1) gauge field\cite{avl1}:
\be
\tilde{H}=-2t\sum_{\la ij\ra }e^{ia_{ij}}\Phi_i^{\dagger}\Phi_j+h.c.+H_{Maxwell}+...
\ee
where one may also have short range vortex interaction terms in general. For spin-half antiferromagnets ({\em i.e} where the original rotor number is $1/2$ per site on average), the vortices will themselves see a background $\pi$-flux 
on each plaquette.

%Usually the vortices will either condense or be gapped. The vortex consensate gives a bosonic Mott insulator, while the gapped bosons give a spin-ordered state (or a superfluid), and the translational symmetry is also likely to be broken due to the finite vortex density (i.e. a vortex Mott insulator at half-filling). 

This system of hard-core bosonic vortices at half-filling could be fractionalized. To explore this 
possibility, we fractionalize the vortex operator $\Phi$ into two fermions using the slave-particle formulation:
\be
\Phi_i=\frac{1}{2}\ep^{\al\beta}\psi_{i,\al}\psi_{i,\beta}, \hspace{20pt} N_i=\frac{1}{2}\psi^{\dagger}_{\alpha}\psi_{\alpha}, %\Phi_i^{\dagger}=-\frac{1}{2}\ep^{\al\beta}\psi_{i,\al}^{\dag}\psi_{i,\beta}^{\dag},
\ee
where $N$ denotes the vortex density, and $\alpha, \beta=1,2$ are the pseudo-spin indices, which transform under the internal $SU(2)$ gauge symmetry as $\psi_{\al}\to U_{\al\beta}\psi_{\beta}$. 
The lattice symmetries act on $\psi_{i,\al}$ in the same ways as on $\Phi$ (up to an $SU(2)$ gauge transform). For a spin model, time 
reversal acts on vortices as  $\mathcal{T}: \Phi_i\to \Phi_i $, we have $\mathcal{T}: \psi_{i,\al}\to\psi_{i,\al}$ (again up to a gauge rotation). 
The particle-hole symmetry  (coming from $\pi$ rotation of spins around $x$ axis) transformation acting on the vortex is non-trivial: 
$\mathcal{C}: \Phi_i\to \Phi_i^{\dag} $, which leads to 
$\mathcal{C}: \psi_{i,\al}\to W_{i,\al\beta}\psi^{\dag}_{i,\beta}$ where $W$ is unitary with ${\rm{det}}(W)=-1$.

Our goal is to explore phases in which the fermions $\psi_{1,2}$ are deconfined and gapless. The gaplessness of the fermions should be stable in the sense that it is protected by symmetries. 
 It is instructive to reinterpret the ``fermionized vortex'' theory of  Refs. \onlinecite{avl1,avl2,avl3} using this dual parton construction. It corresponds to putting $\psi_1$ in a Chern-insulator and $\psi_2$ in a gapless Dirac band. However, since time-reversal is broken in such a phase, the gaplessness is unprotected.

Now consider a particular mean field ansatz that meets our need:
\be
\label{ansatz}
H_{mean}=-\sum_{ij}\lp \psi_{i\alpha}^{\dag}u^{\alpha\beta}_{ij}\psi_{j\beta}+h.c.\rp,
\ee
with the hopping matrices $u_{ij}$ given by
\bea
\label{ans}
u_{i,i+\vec{a}_1}&=&u_{i,i+\vec{a}_2}=u_{i,i+\vec{a}_3}=\eta\tau^0+\lam\tau^3, \\
u_{i,i+\vec{a}_1+\vec{a}_2-\vec{a}_3}&=&u_{i,i+\vec{a}_1-\vec{a}_2+\vec{a}_3}=u_{i,i-\vec{a}_1+\vec{a}_2+\vec{a}_3}=\xi\tau^1, \nonumber
\eea
where  $\vec{a}_i$ are the three nearest-neighbor vectors on the honeycomb lattice, $\eta,\lam,\xi$ are all real and $\tau^l$ are Pauli matrices acting on the $SU(2)$ gauge indices. 
It is easy to see that $\langle\psi^{\dagger}_i\tau_{\mu}\psi_i\rangle=0$ on any site $i$ due to the particle-hole and time-reversal symmetries preserved by the mean field band structure. Therefore the mean field ansatz satisfies the gauge constraints on average and no further chemical potential term is needed. 
To determine the remaining gauge structure in the phase described by Eq.~\eqref{ans}, one needs to calculate the $SU(2)$ gauge fluxes of the hopping matrices $u_{ij}$ on various loops, and 
all the fluxes must be invariant under the unbroken gauge group\cite{wen}. It is then straightforward to see that only the $\mathbb{Z}_2$ gauge group $\psi_i\to(-1)^{s_i}\psi_i$ survives. 

The ansatz given in Eq.~\eqref{ans} realizes all the lattice symmetries trivially, and is also manifestly time-reversal invariant. 
Hence $\psi_{\alpha}$ transforms in exacly the same way as $\Phi$.
For charge conjugation $\mathcal{C}$, by inspection one can see that we should choose $\mathcal{C}: \psi_{i,\alpha}\to i(-1)^i\psi^{\dagger}_{i,\alpha}$, where $(-1)^i$ takes opposite values on different sublattices. 
The fermions $\psi$ should also be coupled to the non-compact $U(1)$ gauge field $a_{\mu}$, and from the structure of the ansatz it is clear that the only way to do this consistently is to assign charge-$1/2$ to both $\psi_{1,2}$.

The virtue of the ansatz Eq.~\eqref{ans} is that it supports a gapless band structure protected by symmetries. It is straightforward to show that the band structure is described by four Dirac cones (similar to Graphene) near $\pm\vec{Q}$, and the low energy `mean field' Hamiltonian can be written as
\bea
\label{eff}
H_{eff}(\vec{k})=&&\frac{\sqrt{3}}{2}\lp\eta\tau^0+\lam\tau^3-2\xi\tau^1\rp \nn
&&\ot\lp k_x\s^1\ot v^3-k_y\s^2\ot v^0\rp,
\eea
where $\s^i$ acts on sub-lattice indices and $v^i$ on valley indices.

The symmetry actions on the low energy fermions in the above basis can be worked out through standard procedures:
we have the lattice translation $T_{(1,0)}: \psi\to \exp\lp i\frac{4\pi}{3}\s^0\ot v^3\rp\psi$; $\pi/3$ rotation around the center of an honeycomb plaquette (a site of the original triangular lattice)
$R_{\pi/3}\psi=\s^2\ot v^2e^{-i\frac{\pi}{6}\s^3\ot v^3}\psi$; modified $x$-reflection 
$\tilde{\mathcal{R}}_x=\mathcal{R}_x\mathcal{C}: \psi(k_x,k_y)\to\tau^0\ot\s^0\ot v^1\psi(-k_x,k_y)$ (note that a simple reflection flips vorticity); charge conjugation $\mathcal{C}:\psi(\vec{k})\to\tau^0\ot\s^3\ot v^1\psi^{\dag}(-\vec{k})$; 
time reversal $\mathcal{T}$ ($\psi(k_x,k_y)\to\tau^0\ot\s^0\ot v^1\psi(-k_x,-k_y)$ and complex conjugation).

We can now analyse generally what fermion-bilinear terms are allowed by symmetries in the low-energy theory. It is then straightforward 
to show that Eq.~\eqref{eff} is the most general form of symmetry-allowed low energy hamiltonian of the fermions. In particular, 
a mass term that opens up a fermion gap is not allowed by symmetries. Hence the gaplessness of the fermions are symmetry-protected, at least perturbatively.

The above analysis can also be applied to a physical hard-core boson system on a honeycomb lattice at half-filling. The resulting state is a gapless $Z_2$ fractionalized liquid. The charge-$1/2$ fermions form four Dirac nodes, with a velocity anisotropy in the pseudo-spin space. As we will see below, when we view the theory instead as a vortex theory, the coupling to the $U(1)$ gauge field $a_{\mu}$ removes the velocity anisotropy at low energy.

The low energy Lagrangian with the $a_{\mu}$ field included can be written as
\be
\label{dirac}
\mathcal{L}=\bar{\psi}\lb-i(\gamma^{\mu}+\hat{\gamma}^{\mu})(\p_{\mu}+i\tilde{a}_{\mu})\rb\psi+\frac{1}{2e^2}f_{\mu\nu}^2.
\ee
We have chosen the normalization $\tilde{a}_{\mu}=a_{\mu}/2$, $\eta=1$ and $\bar{\psi}=i\psi^{\dagger}\gamma^0$, where $\gamma^{\mu}=(\tau^0\ot\s^3\ot v^3, \tau^0\ot\s^2\ot v^0, \tau^0\ot\s^1\ot v^3)$, and $\hat{\gamma}^{\mu}=(0,(\lambda\tau^3-2\xi\tau^1)\ot \s^2\ot v^0, (\lambda\tau^3-2\xi\tau^1)\ot\s^1\ot v^3)$. This is not quite Dirac, but after including the fluctuation of the U(1) gauge field, it will renormalize 
to a Dirac theory with emergent Lorentz symmetry. For small $\lambda$ amd $\xi$ and large $N_f$ (here we have $N_f=4$), we have 
to first order
\be
\frac{1}{\lambda}\frac{d\lambda}{dl}=\frac{1}{\xi}\frac{d\xi}{dl}=-\frac{64}{5\pi^2N_f}.
\ee 
Hence they are irrelevant to first order. The calculation is essentially identical to that in Ref. \onlinecite{asl}, where it was shown that the velocity anisotropy in real space was irrelevant (see Appendix \ref{rg} for details). Hence the low energy fixed point is simply the $QED_3$ with four flavors of Dirac fermions. It is believed that for flavor number $N_f$ not too small (greater than certain critical value $N_{f,c}$), the $QED_3$ fixed point is a CFT that is stable against spontaneous chiral symmetry breaking and fermion mass generation. The currently known\cite{tarun} upper-bound for $N_{f,c}$ is $N_{f,c}<6.6$. If the actual value of $N_{f,c}$ is less than four, our theory would describe a stable critical phase, rather than just a fine-tuned critical point. 

One could also consider slightly modifying the system, by changing the flux on each plaquette in the rotor model Eq.~\eqref{rotor} from $\pi$ to $(\pi+2\pi\delta)$. This changes the vortex filling to $(1/2+\delta)$, which is also the filling fraction of the $\psi$ fermions. The same mean field ansatz Eq.~\eqref{ansatz} would then describe small fermi surfaces coupled with the gauge field $a_{\mu}$. As discussed in Ref. \onlinecite{max}, such a theory could describe a stable phase. However, we will not study this phase in detail since the modified system is harder to realize.
%({\bf Will an XY spin model with Dzyaloshinski-Moriya interaction not do it? This is quite realistic}). 
Since our purpose is mainly to illustrate the new formalism, the $QED_3$ fixed point theory is enough to convey the message.

The critical phase thus obtained has symmetries that are absent in the microscopic model, but emerge at low energy. These include the Lorentz invariance and the $SU(4)$ flavor symmetry. The $SU(4)$ group is generated by $\{\tau^0\ot\s^0\ot v^3, \tau^0\ot\s^2\ot v^1,\tau^0\ot\s^2\ot v^2,\tau^{1,2,3}\ot\s^0\ot v^0\}$ 
and their tensor products, which gives 15 generators in total, denoted by $T^a$, and by construction we have $[T^a,\gamma^{\mu}]=0$.

%This might give an example of ``quantum order'' even beyond the projective symmetry group (PSG) paradigm (\cite{8}). It might be useful to characterize such phases with emergent symmetries, i.e. symmetries that are not present in general but only at stable fixed points controling the phases.

%\subsection{Monopoles}

\section{Physical properties}
Now we look at particular features of the specific gapless vortex liquid state constructed above by considering physical observables. As a critical theory, we expect many of the physical observables will have algebraic correlation functions, and the exponents can be calculated using the CFT description. 
A notable exception, however, is the in-plane spin-spin correlations. A spin-$1$ excitation $S^{\pm}$ is represented as the composite of the vison $s$ seen by the half-vortices $\psi_{\alpha}$ and a half-monopole in $a_{\mu}$. Since the vison $s$ is assumed to be gapped, we expect $S^{\pm}$ to be also gapped, and the in-plane spin-spin correlations $\la S^+S^-\ra $ will thus be short-ranged. 

The out-of-plane spin-spin correlation functions $\la S^zS^z\ra $, on the contrary, decays algebraically. In fact, since $S^z$ is conserved in the CFT with the corresponding current represented as $j\sim da$, its scaling dimension must be $h_j=2$. We therefore have an interesting state with gapped $S^{\pm}$ but critical $S^z$. 
In fact, the rich symmetry structure of our theory gives many other conserved currents which all have scaling dimension $h_j=2$. These include the vorticity $J_{\mu}=-i\bar{\psi}\gamma_{\mu}\psi$ and the $SU(4)$ flavor current $J^a_{\mu}=-i\bar{\psi}\gamma_{\mu}T^a\psi$.

%We can now consider correlations of physical observables in this theory. Since $QED_3$ is a CFT, most of the physical observables 
%will decay algebraically, except for the in-plane spin-spin correlation, which we expect to be short-ranged. This can be seen directly in 
%our theory: a single-spin excitation corresponds to the composition of a half-monopole in $QED_3$ and a $Z_2$ vison, and since the vison is gapped, 
%the single-spin excitation is also gapped.

The more interesting observables are nematic (spin-$2$) order parameters like $(S^+)^2$. In the dual picture these nematic operators are 
represented as monopoles in $QED_3$. There are four flavors of Dirac fermions and  each of them gives a zero-mode in the presence of $\pm 2\pi$ flux of $\tilde{a}_{\mu}$. A gauge-invariant state created by a monopole event should have half of the zero-modes filled. Hence there are six possible monopoles, obtained by filling two of the four zero-modes. We show in Appendix \ref{mono} that the monopole operators indeed transform in the same way as $(S^{\pm})^2$ at the three low energy momenta $(0,\pm\vec{Q})$.

The scaling dimension of the nematic operators is thus given by that of the monopole operators, which 
can be calculated in the large-$N_f$ limit \cite{monopole,monopole2} (here we have $N_f$=4): $h_n\approx0.265N_f-0.038\approx1.02$. The relatively small scaling dimension reveals the proximity to nematically ordered phases.

To actually go to a nematic phase, the fermions $\psi_{\alpha}$ must acquire a mass gap. Since all the fermion mass terms break some global symmetries, the mass gap must be dynamically generated through spontaneous symmetry breaking, which agrees with the intuition that an ordered state on a frustrated lattice should break some symmetries other than the global $U(1)$. 
Possible mass terms are the flavor $SU(4)$ adjoint $N^a=-i\bar{\psi}T^a\psi$ and scalar $M=-i\bar{\psi}\psi$. It turns out\cite{asl} that $M$ has a relatively large scaling dimension, so the primary instability comes from the $N^a$ terms. 
The scaling dimensions of all the $N^a$ operators (which must be the same due to the $SU(4)$ symmetry) have been calculated\cite{asl} to leading order in $1/N_f$ which gives $h_N\approx2-64/3\pi^2N_f\approx1.46$. 
In particular, these include the coplanar order parameter (spin chirality) $\kappa\sim K_z: \tau^{\mu\neq2}\ot\s^3\ot v^0$, and
the collinear order parameter (bond energy wave) $K_{\pm}: \tau^{\mu\neq2}\ot\s^1\ot (v^1\pm i v^2)$, which are expected to order in usual magnetic phases.

The large number of operators with the same relatively small scaling dimensions gives a clear manifestation of the emergent $SU(4)$ flavor symmetry. Physical observables that transform the same way with $N^a$ under microscopic symmetries will thus have the same scaling dimensions $h_N$. 
It is straightforward to see that eight distinct physical operators are connected by the $SU(4)$ flavor symmetry. %while in the second phase, we have twelve distinct physical operators.
We list all the physical operators in Appendix \ref{ops}.

Finally we mention some of the thermodynamic properties of this state. Clearly the low-$T$ heat capacity will be $C \propto T^2$, and the uniform spin susceptibility (for field coupling to $S^z$) will be $\chi^z \propto T$. The proportionality constants will depend on the (non-universal) Dirac velocity $v$ in a universal way such that the Wilson ratio $\frac{T\chi^z}{C}$ is a universal constant characteristic of the CFT (computable in the $\frac{1}{N_f}$ expansion).  

There is another $QED_3$ fixed point for the theory in Eq.~\eqref{dirac}, by choosing the $\gamma$ matrices differently. We discuss this fixed point in Appendix \ref{afp}. We show that physical observables behave differently in this new fixed point, so it is indeed a distinct phase from the one discussed so far.

\section{Relationship to other states}
We now briefly consider how the gapless quantum vortex liquid state is related to other more familiar phases of the quantum $XY$ magnet. We have already  discussed in Sec. \ref{duality} and later that if the vortex fields $\psi$ acquires a gap then the result is a phase with long range spin-nematic order ({\em i.e} where $(S^+)^2$ is ordered without ordering of $S^+$).  As also discussed in Sec. \ref{duality}, if the $\psi$ pair and condense, the result is a $Z_2$ quantum spin liquid but without fractionalization of the global $U(1)$ quantum number. 

Although being conceptually close to a nematic phase, the gapless vortex liquid can also be found near other conventional states in principle, via a direct phase transition. 
 To make a transition into a simple ordered state in which $\la b\ra \neq 0$, simply condense the vison $s$ seen by $\psi$, then the fermions $\psi$ will be confined and the vortex $\Phi$ will be gapped, which is nothing but an ordinary superfluid. 
The trivial Mott insulator is also accessible through condensing the composite of the fermion half-vortex $\psi$ and the vison $s$ (which is a boson $v\sim\psi s$ due to the mutual semion statistics), which will confine all the fractional particles and make the system gapped.

\section{A parton construction}
\label{parton}

Here we give a parton construction of the phases we discuss. For this purpose we consider a modified system, in which a rotor $b\sim e^{i\phi}$ lives on the site of the triangular lattice, and an auxiliary rotor $\tilde{b}$ lives at the center of each plaquette of the triangular lattice. The auxiliary rotors thus form a honeycomb lattice. 
We further demand the $U(1)$ rotation symmetry to act only on the $b$ rotors, but not on the auxiliary $\tilde{b}$ rotors. In other words we allow terms like $\Delta H\sim h\tilde{b}+h.c.$ in the Hamiltonian for the auxiliary sites. Now consider fractionalizing the auxiliary rotors as
\be
\tilde{b}=\Phi_1\Phi_2,
\ee
where $\Phi_1$ and $\Phi_2$ are bosons coupling to an emergent $U(1)$ gauge field $A_{\mu}$. For the $b$ rotors, we go to the dual pictur in terms of the vortex field $\Phi$ and the non-compact $U(1)$ gauge field $a_{\mu}$ whose flux is the charge of the $U(1)$ symmetry of the $b$ rotors. We then condense the following object:
\be
\la\Phi^{\dagger}_1\Phi\ra\neq0.
\ee
This is equivalent to putting the $b$ rotors and the $\Phi_1$ bosons into the $(001)$-hierarchical quantum hall state\cite{iqhe}. The condensate will Higgs the gauge field $A^-_{\mu}=(A_{\mu}-a_{\mu})/2$ and leaves only one gauge field $A^+_{\mu}=(A_{\mu}+a_{\mu})/2$ un-Higgsed. Since the gauge field $a_{\mu}$ in the vortex picture is non-compact, the un-Higgsed gauge field $A^+_{\mu}$ is also non-compact. Furthermore it is easy to check that $2\pi$-flux of $A^+_{\mu}$ carries $2\pi$-flux of $a_{\mu}$, which carries charge-$1$ under the $U(1)$-$XY$ symmetry.
The final effective theory is thus the same as the dual-vortex theory: the uncondensed boson $\Phi_2$ coupling to the non-compact gauge field $A^+_{\mu}$, where the flux of the gauge field carries $U(1)$ charge. We can now further fractionalize $\Phi_2$ as we did for Eq.~\eqref{vfrac}:
\be
\Phi_2=\psi_1\psi_2,
\ee
and the field theory for the phase we discussed thusfar is recovered.

We should emphasize that even though this is a construction in the usual parton language, it is much more natural to discuss our phase in the dual parton language, where the fractionalization if introduced straightforwardly with no auxiliary degrees of freedom required.

\section{Discussion}
We have described a concrete example of a gapless quantum spin liquid phase as a gapless fractionalized quantum vortex liquid. It is certainly desirable to find some concrete spin Hamiltonians to realize such phases. However this task is very challenging at this point. Instead we have focused on the more tractable phenomenological side: if these phases are indeed realized in some spin systems, what are the interesting features that could clearly distinguish them from the more familiar phases? We addressed this issue for the particular example in this work.

%Can this phase be accessed directly within the usual spinon-gauge field paradigm based on some spinon mean field ansatz?  We point out that since the vortices are fractionalized the short distance physics of our state is not that of fractionalization of the spin but rather that of spin nematic-like pairing. This suggests that the spinon mean field + fluctuation paradigm may not be able to reach the spin liquid state described here. Our construction thus presents a potential alternate class of gapless quantum spin liquids that may be beyond the spinon-gauge field paradigm. 

Clearly the dual parton approach developed here can be used to construct a variety of other gapless quantum vortex liquid states. An interesting example is a state where the fractionalized vortices form a gapless Fermi surface rather than Fermi points. The coupling of the vortices to the non-compact gauge field will lead to a low energy field theory similar to that of a spinon Fermi surface spin liquid\cite{sfs1,sfs2} or the HLR state\cite{hlr} of the half-filled Landau level. Of course as in the Dirac case discussed here the identification of physical operators in terms of the fields of the low energy theory will be different and will lead to different physical properties. 

%  Finally we briefly mention a generalization to $SU(2)$ invariant spin systems. Consider for instance a $2d$ spin-$1$ magnet with, say,  a ferro-nematic ground state characterized by uniform spin quadrupolar order with some director $\hat{n}$. The spin itself is gapped in this state. In addition there is a point topological defect - a disinclination - around which $\hat{n}$ winds by $\pi$.  As in the $XY$ case, the unpaired spin may be taken to be a mutual semion with the disinclination. Thus their bound state is a fermion. If we now disorder the nematic ordering by putting  these fermions in a gapless band we will get a gapless quantum spin liquid.

The states described in this paper should open our eyes to other new possible routes to gapless spin liquid behavior and suggest alternate possibilities for building phenomenologies of existing experimental candidates.

\textbf{Acknowledgement}: We thank F. Wang for helpful discussions. This work was supported by NSF DMR-1305741.  This work was partially supported by a Simons Investigator award from the
Simons Foundation to Senthil Todadri.

\appendix

\section{RG of psuedo-spin velocity anisotropy}
\label{rg}
We can re-write the Lorentz-breaking perturbation as
\bea
\Delta\mathcal{L}&=&-i\lambda\bar{\psi}\tau^3(\gamma^1D_1+\gamma^2D_2)\psi \nonumber \\
&=&-i(\lambda/3)\bar{\psi}\tau^3(-2\gamma^0D_0+\gamma^1D_1+\gamma^2D_2)\psi \nonumber \\
&&-i(2\lambda/3)\bar{\psi}\tau^3\slashed{D}\psi.
\eea
The last term can be absorbed into the Dirac term by re-defining $\tilde{\psi}=(1+2\lambda/3\tau_3)^{1/2}\psi$, simplifying the perturbation (to leading order in $\lambda$) to
\bea
\Delta\mathcal{L}&=&-i(\lambda/3)\bar{\psi}\tau^3(-2\gamma^0D_0+\gamma^1D_1+\gamma^2D_2)\psi \nonumber \\
&=&-i(\lambda/3)\bar{\psi}\tau^3(-\gamma^0D_0+\gamma^1D_1)\psi \nonumber \\
&&-i(\lambda/3)\bar{\psi}\tau^3(-\gamma^0D_0+\gamma^2D_2)\psi.
\eea
The last two terms share the same structure with the velocity anisotropy term examined in Ref. \onlinecite{asl}, from which the leading order RG flow follows directly:
\be
\frac{1}{\lambda}\frac{d\lambda}{dl}=-\frac{64}{5\pi^2N_f}.
\ee

\section{Quantum numbers of monopoles}
\label{mono}

The monopole operators are defined through their operations on the zero-flux ground state:
\bea
\label{monodef}
M^{\dag}_{L/R}|0\ra&=&e^{i\theta_{L/R}}f^{\dag}_{1,R/L,+}f^{\dag}_{2,R/L,+}|DS,+\ra, \nn
M_{L/R}|0\ra&=&e^{i\phi_{L/R}}f^{\dag}_{1,L/R,-}f^{\dag}_{2,L/R,-}|DS,-\ra, \nn
M^{\dag}_{\al\beta,0}|0\ra&=&e^{i\theta_{\al\beta,0}}f^{\dag}_{\al,L,+}f^{\dag}_{\beta,R,+}|DS,+\ra, \nn
M_{\al\beta,0}|0\ra&=&e^{i\phi_{\al\beta,0}}f^{\dag}_{\al,L,-}f^{\dag}_{\beta,R,-}|DS,-\ra, \nn
\eea
where $f^{\dag}_{\al,R/L,\pm}$ occupies the zero-mode coming from psuedo-spin $\alpha$ and valley $R/L$ in $\pm 2\pi$ flux, and $|DS,\pm\ra$ denotes the state with all the negative energy levels filled in $\pm 2\pi$ flux. 
The symmetry properties of the zero-modes $f_{\al,R/L,\pm}$ and the filled negative Dirac sea $|DS,\pm\ra$ can be obtained. %The details are similar to those in Ref.\onlinecite{avl1} and are shown in Appedix \ref{mono}. 
The calculation is almost identical to that in Ref. \onlinecite{avl1}. The only difference is that we have four flavors of Dirac fermions instead of two in Ref.\onlinecite{avl1}, which makes our calculation easier due to the cancellation of the sign ambiguities in Ref. \onlinecite{avl1}.

The filled negative Dirac sea is defined through
\be
\label{ds}
|DS,q\ra=e^{iq\gamma}\Pi_{E<0}c^{\dagger}_{Eq}|vac,q\ra,
\ee
where the background flux is $2\pi q=\pm2\pi$, and $|vac,q\ra$ is the state with all the fermion levels unoccupied.

One can choose the phases in the definition of $|vac,q\ra$ so that
\bea
\mathcal{T}|vac,q\ra&=&|vac,-q\ra, \nn
\tilde{R}_x\mathcal{T}|vac,q\ra&=&|vac,q\ra,
\eea
and choose the phase $\gamma$ in Eq.~\eqref{ds} so that
\be
\mathcal{C}|DS,q\ra=f^{\dag}_{1,R,-q}f^{\dag}_{1,L,-q}f^{\dag}_{2,R,-q}f^{\dag}_{2,L,-q}|DS,-q\ra.
\ee

The rest of the symmetry properties are determined by the filled Dirac sea, and are heavily constrained by the algebraic structure of the symmetry groups. The contributions from a filled Dirac sea with two flavors are calculated in Ref. \onlinecite{avl1}, with some sign ambiguities that cannot be determined from the group structure. Fortunately we have two copies of the Dirac sea that transform identically under all the microscopic symmetries. Hence the sign ambiguities cancel, and the symmetry properties are uniquely determined from the group structure.

One can then show that the symmetry properties of the filled negative Dirac sea are given by
\bea
T_{\delta\vec{r}}|DS,q\ra&=&|DS,q\ra, \nn
R_{\pi/3}|DS,q\ra&=&e^{iq2\pi/3}|DS,q\ra, \nn
\mathcal{C}|DS,q\ra&=&f^{\dag}_{1,R,-q}f^{\dag}_{1,L,-q}f^{\dag}_{2,R,-q}f^{\dag}_{2,L,-q}|DS,-q\ra, \nn
\mathcal{T}|DS,q\ra&=&|DS,-q\ra, \nn
\tilde{R}_x\mathcal{T}|DS,q\ra&=&R_x\mathcal{C}\mathcal{T}|DS,q\ra=|DS,q\ra,
\eea 
where $f^{\dagger}$ fills a zero-mode, and $q=\pm 1$ is the monopole strength. The zero modes in the Coulomb gauge transform as
\bea
T_{\delta\vec{r}}f_{R/L,q}T^{-1}_{\delta\vec{r}}&=&e^{\pm i\vec{Q}\cdot\delta\vec{r}}f_{R/L,q}, \nn
R_{\pi/3}f_{R/L,q}R^{-1}_{\pi/3}&=&ie^{-iq\pi/6}f_{L/R,q}, \nn
\mathcal{C}f_{R/L,q}\mathcal{C}^{-1}&=&f^{\dagger}_{L/R,-q}, \nn
\mathcal{T}f_{R/L,q}\mathcal{T}^{-1}&=&\pm qf_{L/R,-q}, \nn
(\tilde{R}_x\mathcal{T})f_{R/L,q}(\tilde{R}_x\mathcal{T})^{-1}&=&f_{R/L,q}.
\eea

One can then choose the phases in Eq.\eqref{monodef} and define $N=M_{12,0}-M_{21,0}$, 
$L_+=M_{11,0}$, $L_-=M_{22,0}$ and $L_0=M_{12,0}+M_{21,0}$, such that
\bea
\label{monoqn}
T_{\delta \vec{r}}&:& M_{L/R}\to e^{\pm i\vec{Q}\cdot\delta\vec{r}}M_{L/R}, N\to N,\nn
&&L_{\pm,0}\to L_{\pm,0}, \nn
R_{\pi/3}&:& M_{L/R}\to M_{R/L}, N\to N, \nn
&&L_{\pm,0}\to -L_{\mp,0}, \nn
\tilde{R}_x\mathcal{T}&:& M_{L/R}\to M_{L/R}, N\to N,\nn
&& L_{\pm,0}\to L_{\pm,0}, \nn
\mathcal{C}&:& M_{L/R}\to M^{\dag}_{R/L}, N\to N^{\dag},\nn
&& L_{\pm,0}\to L^{\dagger}_{\mp,0}, \nn
\mathcal{T}&:& M_{L/R}\to M^{\dag}_{L/R}, N\to N^{\dag}, \nn
&&L_{\pm}\to L^{\dagger}_{\pm}, L_0\to -L_0^{\dagger}.
\eea

The pseudo-spin SU(2) scalar $N$ and $M_{L/R}$ transform as $(S^{\pm})^2$ at the three low 
energy momenta $(0,\pm\vec{Q})$, as expected. The emergence of the $SU(2)$ vector $L_{\pm,0}$ as another set of spin-$2$ operators reveals the emergent flavor symmetry of the theory. 

\section{Physical observables connected by flavor symmetry}
\label{ops}

The operators corresponding to the flavor $SU(4)$ adjoint $-i\bar{\psi}T^a\psi$ are listed in Table \ref{phase1}.

\begin{table}[tt]
%\begin{center}
\begin{tabular}{|>{\centering\arraybackslash}m{1.3in}|>{\centering\arraybackslash}m{2in}|}
\hline
{\bf Fermion bilinears $\psi^{\dagger}\gamma^0T^a\psi$} &  {\bf Representative physical operators} \\ \hline
$\tau^{\mu\neq2}\ot\s^3\ot v^0$ & $\kappa_p\sim i\sum_{i,j\in p}(s_i^+s_j^--s_j^+s_i^-)$  \\ \hline
$\tau^{\mu\neq2}\ot\s^1\ot (v^1\pm i v^2)$ &  $B^{\pm}_{ij}\sim e^{\pm i\vec{Q}\cdot(\vec{r}_i-\vec{r}_j)}(s_i^+s_j^-+s_j^+s_i^-)$ \\ \hline
$\tau^2\ot\s^3\ot v^0$ &  $\kappa_p\lp\sum_{i\in p}s^z_i\rp$  \\ \hline
$\tau^2\ot\s^1\ot (v^1\pm i v^2)$ &  $B^{\pm}_{ij}\lp s^z_i+s^z_j\rp$  \\ \hline
$\tau^{1,3}\ot\s^3\ot v^3$ &  $N^v_p\lp\sum_{i\in p}s^z_i\rp$  \\ \hline
$\tau^2\ot\s^3\ot v^3$ &  $N^v_p\lp\sum_{i\in p}s^z_i\rp^2$  \\ \hline
\end{tabular}
%\end{center}
\caption{Correspondence between slowly decaying fermion bilinears and microscopic operators in phase 1, where $\kappa_p$ is the spin chirality defined on plaquette $p$, $B^{\pm}_{ij}$ is the bond-energy wave operator defined at the Brillouin zone coner $\pm\vec{Q}$, $s_z$ is the $z$-component of the physical spin, and $N^v_p$ is the vorticity on plaquette $p$. }
\label{phase1}
\end{table}%

\section{Another fixed point}
\label{afp}

There is another $QED_3$ fixed point for the theory in Eq.~\eqref{dirac}, by choosing the $\gamma$ matrices differently. For example one can take 
$\gamma^{\mu}=(\tau^0\ot\s^3\ot v^3, \tau^1\ot\s^2\ot v^0, \tau^1\ot\s^1\ot v^3)$, by the same argument we can show that perturbations 
like $\hat{\gamma}^{\mu}=(0,\eta\tau^0\ot \s^2\ot v^0, \eta\tau^0\ot\s^1\ot v^3)$ are irrelevant. This theory still has an SU(4) 
flavor symmetry, generated by 
$\{\tau^{0,1}\ot\s^0\ot v^3, \tau^{0,1}\ot\s^2\ot v^{1,2}, \tau^{2,3}\ot\s^1\ot v^{1,2}, \tau^{2,3}\ot\s^3\ot v^{3,0}, \tau^1\ot\s^0\ot v^0\}$. 
However, since the group structure of the total symmetry (microscopic, Lorentz and flavor) is now different from the previous theory, 
we expect these two theories to be physically distinct, seperated by a critical point at $\eta=\lambda$, where the velocity of one 
psuedospin component vanishes and the band structure changes drastically, although the 
microscopic symmetries are realized in exactly the same way. It is interesting to note that these phases are distinct solely by emergent symmetries.

The operators connected by the emergent $SU(4)$ flavor symmetry is listed in Table \ref{phase2}, which is clearly distinct from the list given in Table \ref{phase1}. Therefore the new fixed point is indeed qualitatively distinct from the phase discussed in the main text.

 \begin{table}[tt]
%\begin{center}
\begin{tabular}{|>{\centering\arraybackslash}m{1.3in}|>{\centering\arraybackslash}m{2in}|}
\hline
{\bf Fermion bilinears $\psi^{\dagger}\gamma^0T^a\psi$} &  {\bf Representative physical operators} \\ \hline
$\tau^{0,1}\ot\s^3\ot v^0$, $\tau^2\ot\s^0\ot v^3$ & $\kappa_p\sim i\sum_{i,j\in p}(s_i^+s_j^--s_j^+s_i^-)$  \\ \hline
$\tau^{0,1}\ot\s^1\ot (v^1\pm i v^2)$, $\tau^2\ot\s^2\ot (v^1\pm i v^2)$ &  $B^{\pm}_{ij}\sim e^{\pm i\vec{Q}\cdot(\vec{r}_i-\vec{r}_j)}(s_i^+s_j^-+s_j^+s_i^-)$ \\ \hline
$\tau^3\ot\s^0\ot v^3$ &  $\kappa_p\lp\sum_{i\in p}s^z_i\rp$  \\ \hline
$\tau^3\ot\s^2\ot (v^1\pm i v^2)$ &  $B^{\pm}_{ij}\lp s^z_i+s^z_j\rp$  \\ \hline
$\tau^1\ot\s^3\ot v^3$ &  $N^v_p\lp\sum_{i\in p}s^z_i\rp$  \\ \hline
$\tau^3\ot\s^0\ot v^0$ &  $\lp s_{1,i}^+s_{2,i}^--s_{2,i}^+s_{1,i}^-\rp/2i$  \\ \hline
\end{tabular}
%\end{center}
\caption{Correspondence between slowly decaying fermion bilinears and microscopic operators in the new fixed point. To find a simple correspondence of the last one, we can imagine having two species of spins on each site $s_1,s_2$ and then couple them in symmetric ways.}
\label{phase2}
\end{table}%


\begin{thebibliography}{99}
 
 \bibitem{wen}
  X. G. Wen, Quantum Field Theory of Many-Body Systems, Oxford (2004).

 \bibitem{kanoda1}  Y. Kurosaki, Y. Shimizu, K. Miyagawa, K. Kanoda, and G. Saito,
Phys. Rev. Lett. 95, 177001 (2005).

 
 
 \bibitem{kato}
 T. Itou, A. Oyamada, S. Maegawa, M. Tamura and R. Kato, Phys. Rev. B 77, 104413 (2008).
 
 \bibitem{takagi}
 Yoshihiko Okamoto, Minoru Nohara, Hiroko Aruga-Katori, and Hidenori Takagi, Phys. Rev. Lett. 99, 137207 (2007).
 
  \bibitem{sfs1}
 O.I. Motrunich, Phys. Rev. B 72, 045105 (2005).

 \bibitem{sfs2}
 S.-S. Lee and P. A. Lee, Phys. Rev. Lett. 95, 036403 (2005).

 
 \bibitem{matthewnumerics}
 Hong-Chen Jiang, Matthew S. Block, Ryan V. Mishmash, James R. Garrison, D. N. Sheng, Olexei I. Motrunich and Matthew P. A. Fisher, Nature 493, 39-44 (2013).
 
 \bibitem{yslee}  J.S. Helton, K. Matan, M.P. Shores, E.A. Nytko,
B.M. Bartlett, Y. Yoshida, Y. Takano, Y. Qiu, J.-H. Chung, D.G.
Nocera,
Phys. Rev. Lett. 98, 107204 (2007).

\bibitem{hiroi}  
Satoshi Yamashita, Tomoya Moriura, Yasuhiro Nakazawa, Hiroyuki Yoshida, Yoshihiko Okamoto, Zenji Hiroi, J. Phys. Soc. Jpn., 79, 083710 (2010). 

\bibitem{balicas}J. G. Cheng, G. Li, L. Balicas, J. S. Zhou, J. B. Goodenough, Cenke Xu, and H. D. Zhou
Phys. Rev. Lett. 107, 197204 (2011). 
 

 \bibitem{dmrg}
 Simeng Yan, David A. Huse and Steven R. White, Science 332, 1173 (2011).
 
 \bibitem{han}
 Tian-Heng Han, Joel S. Helton, Shaoyan Chu, Daniel G. Nocera, Jose A. Rodriguez-Rivera, Collin Broholm and Young S. Lee, Nature 492, 406-410 (2012).

 \bibitem{avl1}
  J. Alicea, O. I. Motrunich, M. Hermele, and M. P. A. Fisher, Phys. Rev. B \textbf{72}, 064407 (2005).
 
 \bibitem{avl2}
  J. Alicea, O. I. Motrunich, M. P. A. Fisher, Phys. Rev. Lett. \textbf{95}, 247203 (2005).

 \bibitem{avl3}
  J. Alicea, O. I. Motrunich, M. P. A. Fisher, Phys. Rev. B \textbf{73}, 174430 (2006).

 \bibitem{topo}
  T. Senthil and M. P. A. Fisher, Phys. Rev. B \textbf{74}, 064405 (2006).

  \bibitem{avts12}
  A. Vishwanath, T. Senthil,  Phys. Rev. X 3, 011016 (2013). 

  \bibitem{hmodl} 
  C. Wang and T. Senthil, Phys. Rev. B 87, 235122 (2013).
  
  \bibitem{2dvrtxmtl}Victor M. Galitski, G. Refael, Matthew P. A. Fisher, and T. Senthil
Phys. Rev. Lett. 95, 077002 (2005). 
  
  \bibitem{hermele}Michael Hermele, 
Phys. Rev. B 79, 184429 (2009). 

  \bibitem{dual}
  C. Dasgupta and B. I. Halperin, Phys. Rev. Lett. 47, 1556 (1981), Matthew P. A. Fisher and D. H. Lee, Phys. Rev. B 39, 2756 (1989). 

 \bibitem{asl}
  M. Hermele, T. Senthil, and M. P. A. Fisher, Phys. Rev. B \textbf{72}, 104404 (2005).

 \bibitem{tarun}
 Tarun Grover, Phys. Rev. Lett. 112, 151601 (2014).
 
 \bibitem{max}
 Max A. Metlitski, David F. Mross, Subir Sachdev and T. Senthil, arXiv:1403.3694.
 
 \bibitem{monopole}
  V. Borokhov, A. Kapustin, and X. Wu, JHEP. 12, 044 (2002).

 \bibitem{monopole2}
 Silviu S. Pufu, Phys. Rev. D 89, 065016 (2014).

 \bibitem{iqhe}
 T. Senthil and Michael Levin, Phys. Rev. Lett. 110, 046801 (2013).
 

 \bibitem{hlr}
 B. I. Halperin, Patrick A. Lee and Nicholas Read, Phys. Rev. B 47, 7312 (1993).
\end{thebibliography}
\end{document}